# Daemons, the "Troitsk anomaly" in tritium beta spectrum, and the KATRIN experiment

E.M.Drobyshevski

*Ioffe Physico-Technical Institute, Russian Academy of Sciences, 194021 St.Petersburg, Russia*
*E-mail: emdrob@mail.ioffe.ru*

**Abstract**

The "Troitsk anomaly" is a strange bump at the end of the tritium β spectrum observed in electron antineutrino mass measurements. It reveals a half-year variation. The nature of the anomaly can be accounted for by interaction of negative DArk Electric Matter Objects - daemons detected by us with the Nb-containing superconducting coils of the gaseous tritium source of the setup. In crossing the system, the daemons, whose flux varies with a half-year period, drag away Nb-containing clusters and excite their nuclei and electronic shells to produce emission of five Auger electron lines within $E$ = 18566-18572 eV. This yields an estimate for the daemon flux $\sim 10^{-7}$-$10^{-6}$ cm$^{-2}$s$^{-1}$. It is shown that in the KATRIN system (KArlsruhe TRItium Neutrino experiment) presently under construction the ratio of the daemon-stimulated Auger bump signal to the useful β-signal should be larger than that in the Troitsk experiment. The potential of using slightly modified systems at Troitsk and Karlsruhe to study the daemons is pointed out.

## 1. Introduction

A question raised invariably in discussions with colleagues of our experiments [1-3] on detection of daemons (DArk Electric Matter Objects, tentatively identified with Planckian elementary black holes or similar objects carrying a negative electric charge of up to $Ze \approx (hc/8)^{1/2} \approx 10e$ [4, 5]) is why in other, much more sophisticated experiments, which deal, true enough, with a search for other particles and effects, no signs of daemon existence have thus far been revealed. And while any nuclear physics experimentalist admits to having observed numerous unidentified and unexplained events, such an answer can hardly leave one satisfied.

Attempts at untangling the subtleties of somebody else's experiment is a risky and extremely unrewarding task. Indeed, one could hardly find a person who could feel at ease both in theory and experimentation in the area of physics of elementary particles, atomic and nuclear physics, in fine points of superconductor technology, in astronomy and celestial mechanics, and what not. Those who are directly involved in this experiment will immediately locate inaccuracies in such an attempt. Nevertheless, we are going to show in what follows, necessarily in a superficial way, and using a particular example of very rigorous tritium experiments aimed at measuring the upper mass limit of the electron antineutrino, which have revealed at a high confidence level a heretofore unknown anomaly in the β spectrum, that these results may be caused by a flux of daemons captured by the Solar system into helio- and geocentric orbits. Furthermore, these experiments permit one, in their turn, to refine the daemon flux and, in this way, the efficiency of the detector employed by us, and possibly the charge of the daemons responsible for the anomaly.

We hope that the lack of our competence in some specific areas will not be so large as to cast doubt on the fairly obvious conclusions of this analysis.

## 2. On measurement of the neutrino mass from the tritium β decay spectrum

The neutrino mass is a key problem on the way to solution of many issues in cosmology, physics of elementary particles, and astrophysics.

A direct method to measure the mass of the electron antineutrino $m_\upsilon$ would appear to study the continuous β spectrum of radioactive nuclear decay, particularly of tritium $^3T = {^3}H \rightarrow He^3 + e^- + \upsilon$, near the end-point electron energy $E_0 = (M_T - M_{3He} - e^-)c^2 = 18574\text{-}18590$ eV (see, e.g., [6-8]). The idea underlying the measurement of $m_\upsilon$ is actually fairly obvious; indeed, if the spectrum ends at a certain $E_{eff} < E_0$, it is the difference $E_0 - E_{eff} = m_\upsilon c^2$ that determines the value of $m_\upsilon$.

Experiments in this direction have been pursued by many groups during the recent years (see [7-11] and references therein). They turned out far from simple; indeed, one had to accumulate reliable statistics and ensure stable operation with a potential difference of ~1 V at a level of 20 kV and a count rate of ~1 mHz. Only two groups, at Troitsk and Mainz, have reached a certain progress on this way. They yield for the upper limit, respectively, $m_\upsilon < 2.5$ eV [12] and $m_\upsilon < 2.2$ eV [13] with a 95% C.L. Both groups used β spectrometers in which the decay β particles propagate from the tritium source through an electrostatic analyzer along the magnetic lines of force, without contacting the walls of the system and the electrodes. The magnetic field ratio at the entrance to the analyzer ($H_{max}$) and in the electrode area ($H_{min}$) was as high ≈(5-8)×10³, which accounts for the corresponding resolution $\Delta E = E(H_{min}/H_{max})$ [7] (here $E$ is the electron energy). An essential difference, in our opinion, was that the Mainz group used as the β source a 40-nm-thick film of frozen tritium 17 mm in diameter, which was located at ~1.5 m from the analyzer, whereas the Troitsk scientists employed a gaseous source $L = 3$ m long and 5 cm in diameter, whose exit was spaced ~4 more meters from the analyzer. In both cases, the magnetic fields guiding β electrons to the analyzer were generated by superconducting coils made of Nb-containing compounds. (In the Livermore experiment [11], one also made use of a gaseous source with $L = 5$ m and 3 cm in diameter, which was mounted 1.5 m away from a Tretyakov-type analyzer with a toroidal magnetic field. The magnetic fields guiding the β particles to the spectrometer were likewise created by superconducting coils (5 coils)).

It is believed [13] that the experimental systems at Mainz and Troitsk have practically exhausted their potential. Further progress may lie in increasing the size of the β spectrometer to 10 m in diameter to improve its sensitivity by an order of magnitude (to ~0.2 eV), for which purpose many groups have to join efforts (the KATRIN experiment [14]). One cannot be certain, however, that straightforward increase in size and energy resolution, with the main principles of operation of the system left unchanged, would provide the desired possibility of measuring sub-eV values of $m_\upsilon$.

## 3. The "Troitsk anomaly"

The point is that measurements of the tritium β spectrum at a level of $m_\upsilon < 10$ eV revealed an excess (a bump or step) in the number of events near $E_0$ (but for $E < E_0$), which resulted in negative values of $m_\upsilon$ [7,8,10,11]. Similar bumps were found to exist also in the two most persistent experiments, at Mainz and Troitsk, which were not discontinued up to 1999. Moreover, a careful analysis of the step anomaly by the Troitsk group in 1994-1999 revealed a half-year periodicity in the position of this step $E_{step}$ relative to $E_0$ in the differential β spectrum with maxima of $E_0 - E_{step}$ in May-June and November-December [12,15,16]. The excess in the number of events (the bump's amplitude) in the region of the anomaly is fairly large, it amounts to a few mHz for a background level of ~20 mHz. In the beginning, the Troitsk results had seemed to find confirmation by the experiment at Mainz [15], but a design refinement of the tritium film detector (removal of the roughness of its wet surface by lowering properly the temperature) permitted cutting the pulse count rate in the vicinity of $E_0$ down to a level indistinguishable from background. It is in this way that one succeeded in lowering the limit to $m_\upsilon < 2.2$ eV (C.L. = 95%) [13]. Nevertheless, the Troitsk effect may hardly be regarded a closed problem; indeed, the confidence level of approximating the step's displacement $E_0 - E_{step}$ or its amplitude by a sine curve with $P = 0.5$ yr is fairly high, for $E_0 - E_{step}$ the correlation coefficient $r = 0.77$ at C.L. > 99.9%, and for the step amplitude $r = 0.65$, C.L. ≈ 99% (so that $E_0 - E_{step}$ and the amplitude are correlated with $r = 0.58$ at C.L. ≈ 98% (our treatment of 16(15) data points from the report of Lobashev [16]).

The issue consists in understanding the nature of the phenomenon.

The half-year periodicity may imply crossing by the Earth of a shell or ring surrounding the Sun. Associating its composition with components capable of influencing in some way the tritium β decay, one may assume, as this was done by Lobashev [15] with a reference to [17,18], that this shell consists of low-energy degenerate relic neutrinos. Their absorption by tritium nuclei brings about neutrinoless β decay of the latter involving emission of electrons with $E \approx E_0$. This assumption would require, however, a neutrino concentration of $\sim 10^{15}$ cm$^{-3}$, which exceeds standard estimates by 13 orders of magnitude [15] and bears a clear signature of an *ad hoc* hypothesis (let alone the unexplained practical absence of the effect at Mainz).

In addition to the periodic step at $E \to E_0$, one could mention also observation of such features in the Troitsk experiment as (*i*) an excess of the count rate over the expected level for $E$ < 18300-18400 eV, as well as (*ii*) a bunch-type background, which manifests itself in a sporadic arrival (~ once in an hour) during 10-20 s of 10-30 pulses with an energy corresponding to the spectrometer potential. It is believed that the bunch events originate from ionization of the residual gas by the electrons produced by decaying tritium nuclei in the spectrometer itself [7].

## 4. Celestial mechanics evolution scenario of daemons and periodicity of their flux onto the Earth

Our experiments on daemon detection were started in 1996. The first positive results did not appear, however, until March 2000 [1]. It was a peak in the distribution of signals in time, $N(\Delta t)$, which were obtained from two spaced parallel horizontal ZnS(Ag) scintillation screens. The time shift of the peak $\Delta t \approx 30$ μs corresponded to the velocity of the object motion $V_{daem} \approx$ 10-15 km/s. Its confidence level was as high as 2.8$\sigma$ (by the present time, it has reached 3.8$\sigma$; for more details, see also [2,3]). The amplitude of the peak yielded for the flux $f_\oplus \sim 10^{-9}$ cm$^{-2}$s$^{-1}$.

One more year of hard work, which helped us to get a better understanding of our seemingly very simple setup, had to pass before we finally came to the conclusion that the poor month-to-month reproducibility of the results stems from a seasonal variation of the daemon flux. We have to admit that our understanding of daemon interaction with the detector is still very far from adequate. The only thing that is clear is that we still do not control some important parameters, with the result that the detector efficiency is far from unity. Roughly speaking, it is determined primarily by the ratio of the path passed by a daemon in active state, where the total charge of the daemon and of the remainder of the nucleus it had captured is negative so that it is capable of capturing and exciting another nucleus (in air, and at the daemon velocity of 15 km/s, it is about 0.5 cm [2]), to the path in which the daemon causes proton decays with an average interval of $\sim 10^{-6}$ s, one after another, in the captured nucleus down to $Z_n < Z$ (this path length $\sim 10^1$ cm). Therefore, for two scintillation screens the detector efficiency, as this appears presently [19], is possibly only $\sim 10^{-2}$-$10^{-3}$, so that in actual fact the daemon flux is higher by far than the directly measured value of $10^{-9}$ cm$^{-2}$s$^{-1}$ and may reach $f_\oplus \sim 10^{-7}$-$10^{-6}$ cm$^{-2}$s$^{-1}$.

Many of the features observed in the $N(\Delta t)$ distribution have a low statistical significance and presently elude interpretation. The only point that appears certain is that the confidence 1 - $\alpha$ of $N(\Delta t)$ differing from $N(\Delta t) = const$, calculated from the $\chi^2$ criterion, is in some cases close to unity. Moreover, 1 - $\alpha$ varies with $P = 0.5$ yr and reaches a maximum in February-March and August-September (see Fig.1, as well as [3]).

On the whole, this all finds a ready explanation if we take into account that our experiment was aimed from the outset at detection of daemons accumulated in the Solar system through their capture from the Galactic disk population into strongly elongated and nearly circular heliocentric orbits crossing the Earth's orbit. Whence their flux onto the Earth, by our rough estimates [20], may exceed by 4-5 orders of magnitude the very low primary flux $f_\oplus \sim 3 \times 10^{-12}$ cm$^{-2}$s$^{-1}$, which derives from the original population of the Galactic halo or disk, to reach $f_\oplus \sim 3 \times 10^{-7}$ cm$^{-2}$s$^{-1}$.

The Sun moves relative to the nearest stellar population (and, probably, to the DM population of the Galactic disk, which has a velocity dispersion ~4-20 km/s [21]) with a velocity of 16.5 km/s in the direction $l = 53^\circ$, $b = 25^\circ$ (Galactic coordinates) [22]. Acted upon by gravitational focusing, many of the daemons cross the body of the Sun to be slowed down to the extent where they fall on it back to become captured into strongly elongated orbits with perihelia lying within the Sun. Because within the Sun objects move in the field of a radially varying rather than point mass, these orbits do not close. The direction to their aphelia shifts continuously opposite to the orbital motion of the object, with the result that daemon trajectories trace out a rosette while remaining at the same time confined to a plane containing, on the average, the Sun's motion direction. Because of the resistance of the Sun's material, these orbits contract gradually, and the daemons settle down toward the center to form there a daemon kernel [23]. If, however, in the course of motion along such an elongated orbit a daemon crosses the Earth's gravitational sphere of action, the perihelion of the orbit will leave the Sun with a high probability and the orbit becomes stable until the next daemon crossing of the Earth's sphere of action [20]. The orbit will no longer contract and trace out a rosette.

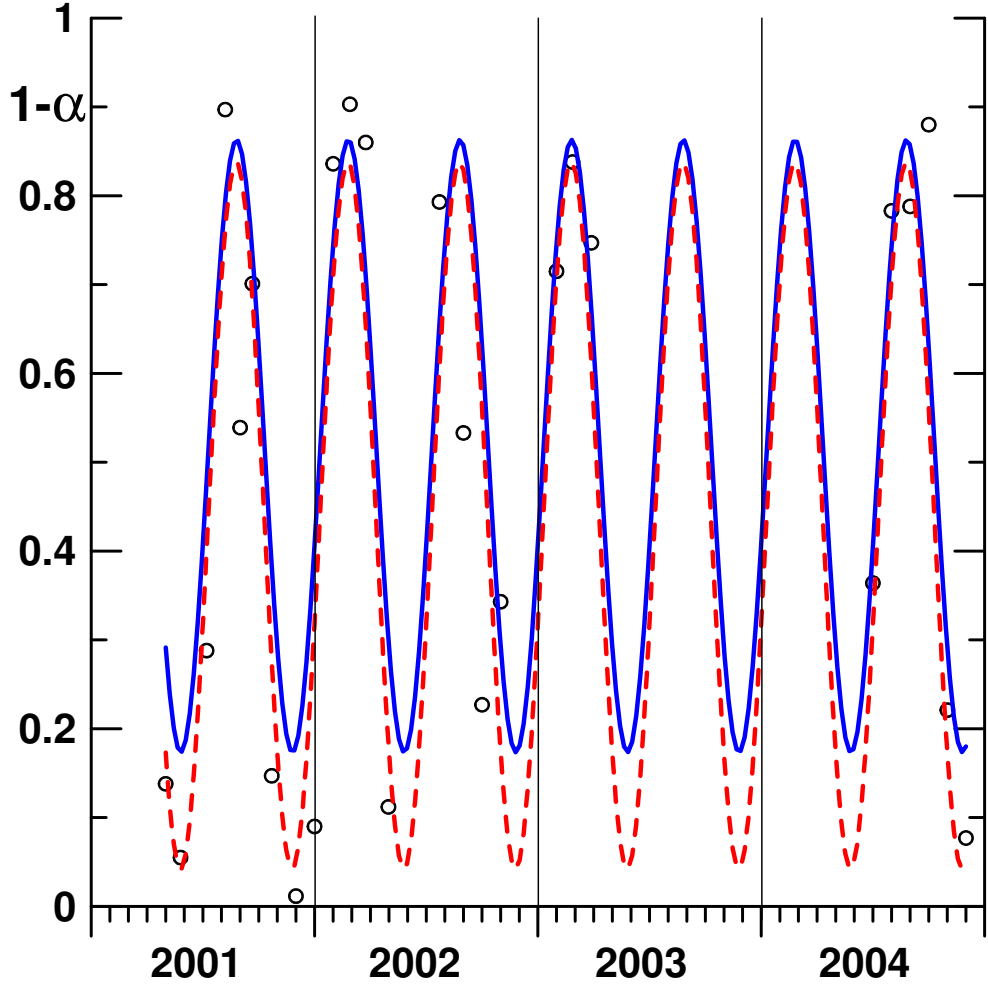

Fig. 1. Seasonal variation of $1 - \alpha$, the extent to which the distribution $N(\Delta t)$ (at $-100 < \Delta t < 100$ µs) deviates from the constant level produced by background events.
(- - - -) Weights of all the points are equal, the correlation coefficient of the sine curve ($P = 0.5$ yr) with the points is $r = 0.87$, its C.L. > 99.9%.
(———) Weights of the points are proportional to $1 - \alpha$, $r = 0.73$, its C.L. = 99.3%.

    As a result of subsequent transits of the daemon through the Earth's sphere of action, its orbit approaches, on the average, ever closer the circular shape to match the orbit of the Earth. In this way the daemons build up in Near-Earth, Almost Circular Heliocentric Orbits (NEACHOs), from which they can fall on the Earth with a velocity $V_{daem} \sim 11\text{-}15$ km/s. Eventually, gravitational perturbations by the Earth are capable to eject some NEACHO objects into the region of the outer planets and, from there, out of the Solar system altogether.

    The plane of the ecliptic is tilted at an angle $\approx 50°$ to the direction of the Sun's motion. It thus follows that the capture into the NEACHO orbits is most efficient in the two areas where the projection of the velocity of the Earth's orbital motion on the direction of the Sun's motion is maximal. These areas are accidentally close to the line of nodes, i.e., the direction to the vernal and autumnal equinoxes (March 21 and September 21). It is in these areas that daemons moving along the strongly elongated, rosette-shaped trajectories spend the longest time in the Earth's

gravitational sphere, which makes them most prone to capture into NEACHOs (objects moving along rosette trajectories in the direction of the Earth's orbital motion should be preferably captured). This suggests that spring and autumn do not provide equal probabilities for detection of daemons on the Earth, and, recalling the 23.5° tilt of the Earth's axis of rotation to the ecliptic, in the Northern and Southern hemispheres in particular. One should therefore expect observation in experimental data, in addition to the half-year periodicity, of a one-year harmonic too.

One may thus conclude that the points where NEACHOs cross the Earth's orbit crowd (while in somewhat different degrees) near the vernal and autumnal equinoxes. In these time intervals, the Earth should meet the maximum fluxes of daemons with $V_{daem} \approx$ 10-15 km/s (but before this, the velocities $V_{daem} >$ 15 km/s and reach as high as ~30-40 km/s [2,3]).

On the other hand, part of the NEACHO daemons, on crossing the Earth, will become captured into Geocentric Earth-Surface Crossing Orbits (GESCOs), which contract gradually (in a time of a few months) to escape under the surface of the Earth as its material slows the daemons down [2,24]. The objects populating these orbits have near-surface velocities of 11 to 0 km/s. Accordingly, their flux becomes shifted in phase relative to that in the NEACHOs. Therefore, the seasonal extrema in $N(\Delta t)$ are also shifted with respect to the equinox periods.

**5. Possible reasons for the "Troitsk anomaly"**

An adequate explanation of the Troitsk anomaly in the tritium β decay spectrum is still lacking. The results of other experiments are fairly ambiguous; the Livermore experiment seemingly confirmed its presence [11], while the improvement of the solid tritium source at Mainz practically removed the anomalies near $E_0$ [13].

On the other hand, the existence of a half-year (and, probably, a year) periodicity of the anomaly, when collated with a similar cyclic pattern in the fall of daemons onto the Earth, which was revealed by us and finds a plausible interpretation within the concepts of celestial mechanics, gives one grounds for suggesting a common origin for the two observations.

One could certainly start with reasons of a fundamental nature and, assuming the style of Stephenson *et al.* [17,18] (see Sec.3), suggest, for example, that daemon-stimulated decay of nucleons in the nuclei of the material of which the walls of the setup are fabricated creates low-energy electron neutrinos (see also, e.g., [25]) to be immediately captured by the neutron of the tritium nucleus, which would culminate in neutrinoless emission of electrons with $E \approx E_0$. Straightforward estimates similar to those proposed in [15] for the relic neutrinos demonstrate, however, that the required neutrino fluxes could not be obtained in this way.

The Troitsk experiment made use of a tritium gas source, $L$ = 3 m long and 5 cm in diameter, which is connected to the spectrometer with a ≈4-m-long channel (in Los Alamos, $L$ = 3.2 m and diameter 3.8 cm [10], in Livermore, $L$ = 5 m and diameter 3 cm [11]). The source and the channel are mounted inside superconducting coils, which generate the magnetic field isolating the β electrons from the walls and leading them to the spectrometer.

The superconducting coils were wound with Nb-based compound wire. It is these materials, on the one hand (and we have been just lucky here), that suit best our interpretation, while on the other, they are most advanced technologically [26, 27]. More specifically, they are $NbTi_n$ (for n ≈ 2) in coil windings of the gas tritium source channels and $Nb_3Sn$ in the coil generating the magnetic field waist at the entrance to the electrostatic analyzer. In the latter case, the Nb wires coated with $Nb_3Sn$ were embedded in $SnCu_n$ bronze (n ≈ 10).

When the $K$ electron is knocked out (its binding energy $E_K$ = 18990 eV), the Nb atom ejects Auger electrons, besides others, with an energy corresponding exactly to the Troitsk bump. The transitions $K–M_2N_2$ occurring with an energy with $E$ = 18568 eV and probability $w_a$ = 1.7, $K–M_2N_3$ with $E$ = 18570 and $w_a$ = 7.8, $K–M_4M_4$ with $E$ = 18566 eV and $w_a$ = 0.002, $K–M_4M_5$ with $E$ = 18569 eV and $w_a$ = 0.01, and $K–M_5M_5$ with $E$ = 18572 eV and $w_a$ = 1.2, five Auger lines altogether, crowd within a narrow region $\Delta E$ = 6 eV (the possible Auger transitions and their probabilities in units of $10^{12}$ s$^{-1}$ were taken from [28, 29], and the binding energies, from [30]; clearly enough, one should exercise utmost care when comparing these values of $E$ with the corresponding values obtained in the Troitsk measurements because of unavoidable systematic differences which may amount to a few eV). The next nearest transitions with a lower energy are transitions $K–M_2N_1$ occurring with an energy $E$ = 18545 eV and probability $w_a$ = 2.7 and $K–M_1N_2$, $E$ = 18480 eV and $w_a$ = 3.1, and $K–M_4N_1$ with a higher energy, $E$ = 18716 eV (>$E_0$ = 18574–18590 eV) and $w_a$ = 0.2. The bump was studied at Troitsk with a resolution of 1 V (and even 0.5 V); indeed, it is near it that $E = E_{eff} \approx E_0$, whereas outside this region (for $E < 18500$ eV and $E > E_0$) the resolution dropped down to 2-5-25 V [7,15].

The process assumed by us consists in the capture by daemons in the superconducting windings of Nb-containing atomic clusters followed by their transport into the magnetic channel leading to the β spectrometer.

It is conceivable that on entering clusters and, subsequently, the nucleus of an atom it contains with liberation of an energy of up to $W = 1.8ZZ_nA_n^{-1/3}$ MeV [31] (for $Z = 10$, this is 162 MeV for Nb with $Z_n = 41$), the daemon (*i*) should shake up and excite electronic shells of the cluster atoms, with subsequent loss of a sizable part (possibly, all) of electrons in some atoms of the cluster (recall the muon physics [29]) and (*ii*) should force the nucleus to eject part of its nucleons and even undergo fission [31], which likewise would entail loss of atomic electrons, including those of neighboring atoms in the cluster. As a result, the cluster would break up, partially or completely, into atoms, some of them strongly ionized, and, possibly, even free nuclei. Electronic shells of atoms are shaken up once every $\sim 10^{-6}$ s by subsequent daemon-stimulated proton decays in a daemon-containing nucleus. Redistribution of electrons left within the cluster and capture of new (refilling) electrons from outside primarily into the outer levels should also give rise to ejection of the monoenergetic electrons of interest to us here. Obviously enough, all the electron component perturbations in the cluster have to increase the chaotic electron background in the system.

At last, a rather remarkable point is that if $Z = 9$, the daemon, after the loss of all atomic electrons from the Sn nucleus it has captured during moving in the $Nb_3Sn$ winding, turns out in an orbit inside the electronic *K* shell of Sn. For the Sn nucleus rotating now around the supermassive daemon, this still corresponds to very high Rydberg levels of the daemon. But when considered relative to the nearest neighbors, the compact daemon−Sn-nucleus complex behaves as a nucleus with $Z_n = 41$, i.e., as a Nb nucleus. As the complex continues to capture new refilling electrons from the outside while the Sn nucleus is sinking to ever deeper levels, it ejects these electrons through the internal conversion or Auger processes over the whole Auger spectrum of such a "niobium", including electrons with energies in the region of the Troitsk bump.

With the present level of our knowledge about daemons, it would be hardly possible to consider all of their conceivable interactions with Nb- or Sn-containing and similar compounds. In this connection it would be certainly of interest to analyze the existence and behavior of daemons with *Z* other than 9 or 10, as well as with a charge multiple of *e*/3.

If the daemon flux at the Earth's surface is $f_\oplus \sim 3 \times 10^{-7}$ cm$^{-2}$s$^{-1}$, then the longitudinal (horizontal) surface area of the Troitsk magnetic pipeline $S = 0.35$ m$^2$ is crossed by one daemon every $\sim 10^3$ s. Moving with a velocity of 10 km/s, it traverses the dia. 3-cm tube in 3 µs and initiates in this time $\sim 3$ nucleon decays in the captured nucleus. If each shakeup of electronic shells in the cluster causes ejection out of it of at least one or two Auger electrons with $E \approx 18570$ eV, this alone could produce a bump with the observed amplitude of 1−3 mHz. For the same flux ($f_\oplus \sim 3 \times 10^{-7}$ cm$^{-2}$s$^{-1}$) created by the daemons captured in March or September from NEACHOs into the slower GESCOs ($V_{daem} \sim 5-3$ km/s), the bump amplitude would increase two or three months later (i.e., in May−June or November−December) by a factor two to three, respectively, bringing us to the figure derived from the Troitsk experiment. Capture and transport of greater clusters by slower daemons could likewise be of significance in this respect. Also, it would be important in measuring the amplitude of the bump near $E_{step}$ to have a possibility to register individual electrons incident with intervals of down to $\sim 1$ µs, the time determined by the mean interval of successive daemon-stimulated nucleon decays in the nucleus, rather than of $\sim 27$ µs, as provided in the neutrino experiments described above [7].

The half-year periodicity of the shift of the step $E_{step}$, i.e., of the bump boundary, relative to $E_0$ (or $E_{eff}$) may be explained in a simplest way by the velocity of the daemon (and of the cluster captured by it), which depends on the season ($V_{daem} = 30-10-5$ km/s), becoming superimposed on the velocity of the ejected Auger electron (for $E = 18570$ eV, $V_e = 80800$ km/s). As a result, the electron energy (for $V_{daem} \approx 10-15$ km/s in March or September) will change exactly by $\Delta E \approx m_e V_e V_{daem} \approx 5$ eV, which will bring about the corresponding broadening of electron Auger lines and the shift of the step. This fits well the amplitude of the shift $E_0 - E_{step} \approx 5$ eV quoted by Lobashev [15,16] (when determining the value of $E_0 - E_{step}$, one should not forget that the quantity $E_0$ itself and Auger line positions are known to within a few eV, see above). In the period preceding March (or September), the velocity with which daemons strike the Earth is in excess of 10−15 km/s, so that the blue wing of the Auger lines of our group with $E > 18560$ eV approaches $E_0$. After March (September), the Earth's surface is crossed primarily by GESCO daemons with $V_{daem} \to 0$ [2,3], with the result that the lines of our Auger group narrow and approach the width corresponding to atoms emitting at rest. The blue boundary of this group, $E_{step}$, shifts away from $E_0$. Everything occurs as observed in the Troitsk experiment. (We have assumed here that electrons are ejected from a moving daemon-containing cluster in arbitrary directions; magnetic field may set a preferred direction, with the result that the Auger line will undergo not only broadening but a shift as a whole as well; this aspect would need, however, an additional analysis.)

Because the longitudinal area of the shorter guiding magnetic pipeline at Mainz is smaller than that at Troitsk, the bump amplitude should also be accordingly smaller, in actual fact, indistinguishable from background (there may be no bump at all if the coil windings are made of a compound containing no Nb atoms).

## 6. Some prospects and the KATRIN experiment

Examination of the results of the experiments on direct determination of neutrino mass by studying the tritium β spectrum near its end-point energy $E_{eff}$ stirs a mixed feeling of wonder and admiration by the skill and persistence of the researchers which they demonstrated in an analysis of the fine effects which were observed with the extremely sophisticated experimental setups.

The nature of the Troitsk anomaly certainly deserves further investigation. As we have seen, one cannot exclude the possibility that it indeed originates from DM objects and that scientists have stumbled here on the first unique manifestations of the existence of daemons. In this case, the Troitsk setup could well be employed, after comparatively minor and even simplifying modifications, in DM studies, including measurements in the region of other Auger lines at $E \neq E_{step}$. These modifications would include shutting off the tritium supply system and the differential pumping system associated with these channels etc., but preservation and, possibly, increase of the longitudinal cross section of the vacuum and magnetic systems of the tritium source, as well as operating the electron gun in continuous mode to provide an additional source of refilling electrons. (Interestingly, the check experiment run without tritium which was described in [32] did not reveal a bump, possibly because of a lack of refilling electrons in the system.)

In our opinion, the large-scale KATRIN system which is presently under construction [14] propels intense interest. Indeed, the tritium gas source (length $L$ = 10 m, diameter $\varnothing$ = 9 cm, $H$ = 6 T), together with the magnetic channel guiding β electrons to the spectrometer, is 35 m long. This provides a longitudinal area of about 3.15 m$^2$, which exceeds the corresponding area in the Troitsk setup by about a factor nine. And if the windings of the guiding superconducting coils are made of Nb-containing compounds with the same density per unit length and filling coefficient (which is apparently indeed the case, because the magnetic fields are practically the same, 5 T and 6 T), the bump amplitude should be likewise ~9 times higher. On the other hand, the ratio of the amounts of tritium contained in the gaseous sources of KATRIN (tritium pressure $p$ = 0.004 mbar at $T$ = 30 K) and Troitsk (partial tritium pressure $p$ = 0.007 mbar at $T$ = 26−28 K) is $(L\varnothing^2 p/T)_{KAT}/(L\varnothing^2 p/T)_{Troitsk}$ = 5.35. Whence it follows that the amplitude ratio of the Troitsk bump to the signal carrying information on the value of $m_\upsilon$ should increase in the KATRIN system ~9/5.35 = 1.7 times (and, possibly, a factor 3 more, if we take into account the increase of the path in which the daemons carrying Nb-containing clusters and crossing the gas source or the channel with guiding magnetic field disintegrate protons in these clusters with the resultant Auger electron emission). Thus, the KATRIN system (without tritium) can be used in DM studies with a still higher efficiency than the Troitsk setup. And because one cannot be absolutely sure that $m_\upsilon \neq 0$, one cannot exclude the possibility that it is the investigation of the properties of daemons that will turn out the main and very important outcome of the KATRIN experiment. The time has probably come when one should seriously think about designing a dedicated setup based on the Troitsk effect and devoted exclusively to the investigation of daemons.

Bearing in mind, however, $m_\upsilon$ measurements, one could suggest the following recommendations for the KATRIN design:

(1) The superconducting coils should be made of materials with an atomic number $Z_n < 41$, e.g. of $V_3Ga$ [27] (so that the $K$ electron binding energy will be <18 keV);

(2) The volume and longitudinal area of the tritium source and of the channels guiding the β electrons to the spectrometer should, if possible, be reduced (this is partially done in the version with a solid tritium source).

**7. Main conclusions**

Thus, the concept of the existence and properties of daemons based primarily on our experiments [1−3] offers a possibility:

(1) *To explain* the Troitsk anomaly as due to emission of Auger electrons in the excitation of atoms in the Nb-containing clusters, which are captured by daemons in the superconducting coils and transported by them into the tritium source channels, in particular
    (a) the position of the step in the β spectrum;
    (b) the seasonal variation of its position, including the phase of the variation;
    (c) the seasonal variation of its amplitude, including its phase;
    (d) practical absence of the Troitsk anomaly in the experiments with solid-tritium β source at Mainz.

(2) *To refine* from the results of the Troitsk experiment:
    (a) the daemon flux ($f_\oplus \sim 10^{-7}$–$10^{-6}$ cm$^{-2}$s$^{-1}$), a figure in agreement with the fairly low efficiency of our detector, which is triggered only by a part of the daemons crossing it; this stresses the need for further refinement of the parameters of the Earth's daemon kernel [24];
    (b) possibly, the charge of the daemon ($Ze = 9e$, which will require additional studies).

(3) *To predict*:
    (a) the presence of bumps (electronic Auger lines) in other locations in the β spectrum of the Troitsk setup, provided its design is left unchanged;
    (b) the presence of bumps, of a still larger *relative* magnitude than in Troitsk, in the KATRIN experiment in the case of the gaseous tritium source.

(4) *To suggest recommendations* for future experiments on neutrino mass measurements from the spectrum of tritium β decay:
    (a) to bring to a minimum the volume and longitudinal area of the gaseous tritium source channels;
    (b) to use in the source coils compounds with atomic number $Z_n < 41$ (the *K* electron binding energy in them should be $E_K < E_0 = 18574$–$18590$ eV).

(5) *To point out* that systems making use of the Troitsk anomaly effect, i.e., Auger effect in intermediate and heavy elements, can be employed to advantage in the detection and study of the properties of daemons. Their potential in this respect should be scrutinized.

(6) *To express confidence* that if researchers engaged in other experiments dealing with nuclear phenomena, cosmic rays etc., including detection of other DM candidates, paid more attention to observations defying standard explanation, as this was done at Troitsk, many of these experiments would have revealed indications of the existence of the Dark Electric Matter Objects.

**Acknowledgements**

The author feels indebted to Academician V.M.Lobashev for his kindly providing information on the materials used in the Troitsk setup. Thanks are due also to Academician V.A.Rubakov for drawing my attention to the half-year periodicity of the Troitsk anomaly.